\documentclass[aps,prl,preprint,superscriptaddress,showpacs]{revtex4}

\def\SomeSpaceIfPreprint{\quad\\[2cm] \Large}
\def\bellelogo{\vbox to 16mm{
               \vss\hbox{\resizebox{!}{3cm}{
               \includegraphics{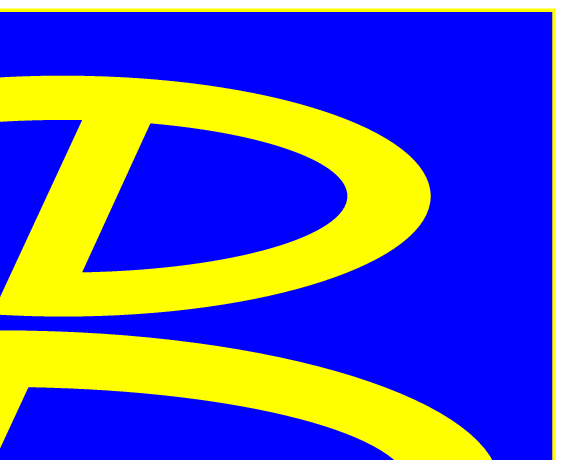}}}}\vspace{0cm}}
\def\preprintA{\hbox{\hfil BELLE-CONF-0414}}
\def\preprintB{\hbox{\hfil ICHEP04 11-0659 }}

\usepackage{graphicx}

\draft 

\def\BB   {B\bar{B}}
\def\qqbar{q\bar{q}}

\def\BtoXsgamma{B\to X_s\gamma}
\def\Xs{X_s}
\def\BtoKll{B\to K\ell^+\ell^-}
\def\BtoKstarll{B\to K^*\ell^+\ell^-}

\def\BtoXsll{B\to\Xs\ell^+\ell^-}
\def\BtoXsee{B\to\Xs e^+e^-}
\def\BtoXsmumu{B\to\Xs\mu^+\mu^-}
\def\BtoXsemu{B\to\Xs e^\pm\mu^\mp}
\def\BtoJpsiX{B\to J/\psi X}

\def\GeV{{\rm~GeV}}
\def\GeVc{{\rm~GeV}/c}
\def\GeVcc{{\rm~GeV}/c^2}
\def\MeV{{\rm~MeV}}

\def\MeVcc{{\rm~MeV}/c^2}
\def\cm{{\rm~cm}}
\def\fbinv{{\rm~fb}^{-1}}

\def\Mbc{M_{\rm bc}}

\def\DeltaE{\Delta E}

\def\Mll{M_{\ell^+\ell^-}}

\def\Mxs{M_{X_s}}

\def\Br{{\cal B}}
\def\calL{{\cal L}}

\def\Fsfw{{\cal F}_{\rm SFW}}
\def\Fmiss{{\cal F}_{\rm miss}}

\def\EffKaon{90\%}
\def\FakeKaon{6\%}

\def\EffBtoXsee{   2.59\,\pm0.18\,^{+0.40}_{-0.37}}
\def\EffBtoXsmumu{ 2.89\,\pm0.14\,^{+0.46}_{-0.43}}
\def\EffBtoXsll{   2.74\,\pm0.13\,^{+0.43}_{-0.40}}

\def\SigBtoXsee{   3.6}
\def\SigBtoXsmumu{ 4.7}
\def\SigBtoXsll{   5.8}

\def\BrBtoXsee{   4.04\,\pm1.30\,^{+0.80}_{-0.76}}    
\def\BrBtoXsmumu{ 4.13\,\pm1.05\,^{+0.73}_{-0.69}}  
\def\BrBtoXsll{   4.11\,\pm0.83\,^{+0.74}_{-0.70}}    
\def\BrBtoXsllFull{4.11\,\pm 0.83({\rm stat})\,^{+0.74}_{-0.70}({\rm syst})}

\begin{document}


\bellelogo

\preprint{\vbox{
  \preprintA
  \preprintB
}}

\title{\large
\SomeSpaceIfPreprint
Improved Measurement of the Electroweak Penguin Process            
\boldmath$\BtoXsll$  
}

\affiliation{Aomori University, Aomori}
\affiliation{Budker Institute of Nuclear Physics, Novosibirsk}
\affiliation{Chiba University, Chiba}
\affiliation{Chonnam National University, Kwangju}
\affiliation{Chuo University, Tokyo}
\affiliation{University of Cincinnati, Cincinnati, Ohio 45221}
\affiliation{University of Frankfurt, Frankfurt}
\affiliation{Gyeongsang National University, Chinju}
\affiliation{University of Hawaii, Honolulu, Hawaii 96822}
\affiliation{High Energy Accelerator Research Organization (KEK), Tsukuba}
\affiliation{Hiroshima Institute of Technology, Hiroshima}
\affiliation{Institute of High Energy Physics, Chinese Academy of Sciences, Beijing}
\affiliation{Institute of High Energy Physics, Vienna}
\affiliation{Institute for Theoretical and Experimental Physics, Moscow}
\affiliation{J. Stefan Institute, Ljubljana}
\affiliation{Kanagawa University, Yokohama}
\affiliation{Korea University, Seoul}
\affiliation{Kyoto University, Kyoto}
\affiliation{Kyungpook National University, Taegu}
\affiliation{Swiss Federal Institute of Technology of Lausanne, EPFL, Lausanne}
\affiliation{University of Ljubljana, Ljubljana}
\affiliation{University of Maribor, Maribor}
\affiliation{University of Melbourne, Victoria}
\affiliation{Nagoya University, Nagoya}
\affiliation{Nara Women's University, Nara}
\affiliation{National Central University, Chung-li}
\affiliation{National Kaohsiung Normal University, Kaohsiung}
\affiliation{National United University, Miao Li}
\affiliation{Department of Physics, National Taiwan University, Taipei}
\affiliation{H. Niewodniczanski Institute of Nuclear Physics, Krakow}
\affiliation{Nihon Dental College, Niigata}
\affiliation{Niigata University, Niigata}
\affiliation{Osaka City University, Osaka}
\affiliation{Osaka University, Osaka}
\affiliation{Panjab University, Chandigarh}
\affiliation{Peking University, Beijing}
\affiliation{Princeton University, Princeton, New Jersey 08545}
\affiliation{RIKEN BNL Research Center, Upton, New York 11973}
\affiliation{Saga University, Saga}
\affiliation{University of Science and Technology of China, Hefei}
\affiliation{Seoul National University, Seoul}
\affiliation{Sungkyunkwan University, Suwon}
\affiliation{University of Sydney, Sydney NSW}
\affiliation{Tata Institute of Fundamental Research, Bombay}
\affiliation{Toho University, Funabashi}
\affiliation{Tohoku Gakuin University, Tagajo}
\affiliation{Tohoku University, Sendai}
\affiliation{Department of Physics, University of Tokyo, Tokyo}
\affiliation{Tokyo Institute of Technology, Tokyo}
\affiliation{Tokyo Metropolitan University, Tokyo}
\affiliation{Tokyo University of Agriculture and Technology, Tokyo}
\affiliation{Toyama National College of Maritime Technology, Toyama}
\affiliation{University of Tsukuba, Tsukuba}
\affiliation{Utkal University, Bhubaneswer}
\affiliation{Virginia Polytechnic Institute and State University, Blacksburg, Virginia 24061}
\affiliation{Yonsei University, Seoul}
  \author{K.~Abe}\affiliation{High Energy Accelerator Research Organization (KEK), Tsukuba} 
  \author{K.~Abe}\affiliation{Tohoku Gakuin University, Tagajo} 
  \author{N.~Abe}\affiliation{Tokyo Institute of Technology, Tokyo} 
  \author{I.~Adachi}\affiliation{High Energy Accelerator Research Organization (KEK), Tsukuba} 
  \author{H.~Aihara}\affiliation{Department of Physics, University of Tokyo, Tokyo} 
  \author{M.~Akatsu}\affiliation{Nagoya University, Nagoya} 
  \author{Y.~Asano}\affiliation{University of Tsukuba, Tsukuba} 
  \author{T.~Aso}\affiliation{Toyama National College of Maritime Technology, Toyama} 
  \author{V.~Aulchenko}\affiliation{Budker Institute of Nuclear Physics, Novosibirsk} 
  \author{T.~Aushev}\affiliation{Institute for Theoretical and Experimental Physics, Moscow} 
% ...
\collaboration{The Belle Collaboration}

\date{\today}


\tighten

\begin{abstract}
We present an improved measurement of the branching fraction for the
electroweak penguin process $\BtoXsll$,
where $\ell$ is an electron or a muon
and $\Xs$ is a hadronic system containing an $s$-quark.
The measurement is based on a sample of 
$152 \times 10^{6}$~$\Upsilon(4S) \to \BB$ events
collected with the Belle detector at the KEKB $e^+e^-$ 
asymmetric-energy collider.
The $\Xs$ hadronic system is reconstructed 
from one $K^{\pm}$ or $K^{0}_{s}$ and up to 
four pions, where at most one pion can be neutral.
Summing over both lepton flavors, 
the inclusive branching fraction is measured to be
$\Br(\BtoXsll)=(\BrBtoXsllFull)\times10^{-6}$
for $m(\ell^{+} \ell^{-}) > 0.2\GeVcc$.

\end{abstract}




\maketitle



\section{Introduction}
\label{sec:Introduction}

In the Standard Model (SM), 
the rare decay $\BtoXsll$ proceeds through a 
$b \to s \ell^{+} \ell^{-}$ transition, 
which is forbidden at tree level.
Such a flavor-changing neutral current (FCNC)
process can occur at higher order via electroweak penguin and $W^{+} W^{-}$
box diagrams.
The $b \to s \ell^{+} \ell^{-}$ transition therefore allows 
deeper insight into the effective Hamiltonian that describes FCNC processes
and is sensitive to the effects of non-SM physics that may enter these
loops; see, for example, Refs.~\cite{Ali02,Hurth03}.

Recent SM calculations
of the inclusive $\BtoXsll$ branching fractions predict
${\cal B}(\BtoXsee) = (6.9 \pm 1.0) \times 10^{-6}$
[$(4.2 \pm 0.7) \times 10^{-6}$ 
for $m(\ell^{+} \ell^{-}) > 0.2$ GeV/$c^2$]
and
${\cal B}(\BtoXsmumu) = (4.2 \pm 0.7) \times 10^{-6}$~\cite{Ali02,Ali_ICHEP02}.
Both the Belle and BaBar Collaborations have 
observed exclusive 
$B \to K \ell^{+} \ell^{-}$  and 
$K^{*} \ell^{+} \ell^{-}$
decays~\cite{Belle_Kll,BaBar_Kll,Belle_Kstll,BaBar_Kstll}
and have measured inclusive $\BtoXsll$ decay~\cite{Belle_sll,BaBar_sll}.

In this analysis, we study the inclusive $\BtoXsll$
process by semi-inclusively reconstructing 
the final state from a pair of electrons
or muons and a hadronic system consisting of one $K^{\pm}$ or $K^{0}_{s}$ 
and up to four pions, where at most one pion can be neutral.
This semi-inclusive-reconstruction approach~\cite{CLEO_bsgamma95}
allows approximately 53\% of the full inclusive rate to be reconstructed.
If the fraction of modes containing a $K^0_L$ 
is taken to be equal to that containing a $K^0_S$,
the missing states that remain unaccounted for represent $\sim$30\%
of the total rate.
We require the hadronic mass for the selected final states
to be less than 2.0$\GeVcc$ to reduce combinatorial background. 
We correct for the missing modes and the effect of the 
hadronic mass cut to extract the 
inclusive $\BtoXsll$ decay rate for $m(\ell^{+} \ell^{-}) > 0.2\GeVcc$.

The measurement of inclusive $\BtoXsll$ decay in this paper
updates and supersedes our previous
publication\cite{Belle_sll} described above, which was based 
on a sample of $65 \times 10^{6}$~$\BB$ pairs.
The measurement reported here is currently the most precise.

\section{The Belle Detector and Data Sample}

We use a data sample collected on the $\Upsilon(4S)$ resonance with the Belle
detector at the KEKB $e^+e^-$ asymmetric-energy collider (3.5 GeV on 8
GeV) \cite{bib:kekb-nim}.  This sample comprises
$152 \times 10^{6}$ $B$ meson pairs, 
corresponding to an integrated luminosity of $140.0 \rm{fb}^{-1}$.
%
%
A detailed description of the Belle detector can be found 
elsewhere~\cite{bib:belle-nim}.
A three-layer silicon vertex detector (SVD) and 
a 50-layer central drift chamber (CDC) are used for
tracking and particle identification for charged particles. 
An array of aerogel threshold \v{C}erenkov
counters (ACC) and  time-of-flight scintillation counters (TOF)
are used for the charged particle identification.
An electromagnetic calorimeter comprised of Tl-doped CsI crystals
(ECL) measures the energy of electromagnetic particles
and is also used for electron identification.
These detectors are located inside a superconducting 
solenoid coil that provides a 1.5~T magnetic field.
An iron flux-return located outside of the coil is instrumented
with resistive plate counters to identify muons (KLM).

Particle identification for the final state particles 
$e^{\pm}$, $\mu^{\pm}$, $K^{\pm}$, $K^{0}_{s}$, $\pi^{\pm}$ and $\pi^{0}$
is important for this analysis. 
Electron identification is based on 
the ratio of the cluster energy to the track momentum ($E/p$), 
the specific energy-loss measurement ($dE/dx$) with the CDC, 
the position and shower shape of the cluster in the ECL and 
the response from the ACC.
Muon identification is based on the
hit positions and the depth of penetration into the ECL and KLM.
Electrons and muons are required to have lab-frame momenta
greater than $0.4\GeVc$ and $0.8\GeVc$, respectively. 
For the muon identification, we also apply a kaon veto
to select good muon candidates. 
Bremsstrahlung photons from electrons are recovered by
combining an electron with photons
within a small angular region around the electron direction.
Charged kaon candidates are selected by
using information from the ACC, TOF and CDC.  
The kaon
selection efficiency is $\EffKaon$ with a pion to kaon
mis-identification probability of $\FakeKaon$.
After selecting the electron, muon and charged kaon candidate
tracks, the remaining charged particles are assumed to be 
charged pions.
$K^{0}_{s}$ candidates are reconstructed from pairs 
of oppositely-charged tracks with 
$|m(\pi^{+}\pi^{-}) - m(K^{0}_{s})| < 15\MeVcc$.
We impose additional $K^0_S$ selection criteria based on the
distance and the direction of the $K^0_S$ vertex and the impact parameters
of the daughter tracks.
We require the charged tracks except for those used
in the $K^0_S$ reconstruction to have impact parameters with respect to
the nominal interaction point of less than $1.0\cm$ in the radial
direction and $5.0\cm$ along the beam direction.
Neutral pions are required to have lab-frame energy greater than $400\MeV$, 
photon daughter energies greater than $50\MeV$,
and a $\gamma \gamma$ invariant mass that satisfies 
$|m(\gamma \gamma) - m(\pi^{0})| < 10\MeVcc$.

\section{Analysis overview}
\label{sec:overview}
In this analysis, we reconstruct the inclusive $\BtoXsll$ decays 
with a semi-inclusive-reconstruction technique from a pair of electrons
or muons and one of 18 reconstructed hadronic states.
Here the hadronic system consists of one $K^{\pm}$ or $K^{0}_{s}$ 
and up to four pions (at most one pion can be neutral).
Compared to a fully inclusive approach, this method has the advantage
of having strong kinematical discrimination against background
by using the 
beam-energy constrained mass 
$\Mbc=\sqrt{E_{beam}^2 - p_{B}^2}$
and the energy difference $\DeltaE = E_{B} - E_{beam}$,
where $E_{beam}$ is the beam energy
and $E_{B}$ ($p_{B}$) is the 
reconstructed $B$ meson energy (3-momentum).
All quantities are evaluated in the $e^{+}e^{-}$ center-of-mass system (CM). 

In addition to the discrimination, further background suppression
to reduce the large combinatorial backgrounds is necessary. 
The main contribution to the combinatorial background comes
from semileptonic decays in $\BB$ events.
In these events, $\BtoXsll$ candidates are reconstructed with
the decay products from both $\BB$ mesons.
This background has a significant amount of missing energy
due to the neutrinos from the semileptonic decays.
Another contribution to the combinatorial background comes from
continuum events, which are effectively suppressed 
with event-shape variables.

There are two background sources that can peak in $\Mbc$ and
$\DeltaE$.
The first comes  from
$B \to J/\psi X$ and $B \to \psi(2S) X$ decays with
$J/\psi (\psi(2S)) \to \ell^{+} \ell^{-}$.
This peaking background is efficiently removed with cuts on the 
dilepton mass $m(\ell^{+} \ell^{-})$. The resulting veto sample provides
a large control sample of decays with a signature identical to that
of the signal.
The second comes from $B \to K^{\pm} (K^0_S) n \pi$~$(n > 1)$ 
decays with misidentification of two charged pions as leptons.
We estimate these peaking background contaminations, then
subtract them from the signal yield.

For the $\BtoXsll$ event simulation,
we use EVTGEN~\cite{EVTGEN} for the event generator,
JETSET~\cite{JETSET} to hadronize the system consisting of a
 strange quark and a spectator quark, 
and GEANT~\cite{GEANT3} for the detector simulation.
In the event generation, $\BtoXsll$ events 
are produced with a combination of 
exclusive and inclusive models.
In the hadronic mass region of $m(X_{s}) < 1.1\GeVcc$, 
exclusive $B \to K^{(*)} \ell^{+} \ell^{-}$ decays are generated 
according to Refs.~\cite{Ali02,Ali00},
where the relevant form factors are computed 
using light-cone QCD sum rules. 
In the region $m(X_{s}) > 1.1\GeVcc$, event generation is based
on a non-resonant model following Refs.~\cite{Ali02,Kruger96} 
and the Fermi motion model of Ref.~\cite{Ali79}.

\section{Event selection}
\label{sec:selection}

Events are required to have a well determined primary vertex,
be tagged as multi-hadron events,
and contain 
two electrons (muons) 
having lab-frame momenta greater than 
$0.4\GeVc$ for electrons and $0.8\GeVc$ for muons.
Dilepton candidates are selected for $e^{+} e^{-}$ or
$\mu^{+} \mu^{-}$ pairs. 
Both leptons are required to originate from the same vertex 
and satisfy the requirement $|\Delta z| < 0.015$~cm. 
Here, $\Delta z$ is the distance between the two leptons 
along the beam direction; the $z$-coordinate 
of each lepton is determined at the point of closest approach to the beam axis.

Charmonium backgrounds are reduced by
removing $B$~candidates with a dilepton mass in the ranges
$-0.40\GeVcc < M_{ee(\gamma)}-M_{J/\psi}   < 0.15\GeVcc$,
$-0.25\GeVcc < M_{\mu\mu}    -M_{J/\psi}   < 0.10\GeVcc$,
$-0.25\GeVcc < M_{ee(\gamma)}-M_{\psi(2S)} < 0.10\GeVcc$, and
$-0.15\GeVcc < M_{\mu\mu}    -M_{\psi(2S)} < 0.10\GeVcc$.
If one of the electrons from a $J/\psi$ or $\psi(2S)$ decay erroneously 
picks up a random photon in the Bremsstrahlung-recovery process,
the dilepton mass can increase sufficiently to evade
the above cuts.
Therefore the charmonium veto is applied to the dilepton mass 
before and after Bremsstrahlung recovery.
Using the simulation, we estimate the remaining peaking 
charmonium background to be $1.20 \pm 0.28$ events and
$1.33 \pm 0.21$ events for $e^{+} e^{-}$ modes and $\mu^{+} \mu^{-}$
modes, respectively.

The potential peaking background from $B \to X_{s} \gamma$ decays,
followed by conversion of the photon into an $e^{+} e^{-}$ pair
in the detector material, and $\pi^0$ Dalitz decay
is a concern for the $e^{+} e^{-}$ modes only. 
We remove this background by requiring
$m(e^{+}e^{-}) > 0.2\GeVcc$. 

Using the $\ell^{+} \ell^{-}$ pair,
$B \to X_{s} \ell^{+} \ell^{-}$ candidates are formed 
by adding either a $K^{\pm}$ or a $K^{0}_{s}$ and up to
four pions, but no more than one $\pi^{0}$. 
In this manner, eighteen different hadronic topologies
are considered:
$K^{\pm}$, 
$K^{\pm} \pi^0$, 
$K^{\pm} \pi^{\mp}$, 
$K^{\pm} \pi^{\mp} \pi^0$, 
$K^{\pm} \pi^{\mp} \pi^{\pm}$,
$K^{\pm} \pi^{\mp} \pi^{\pm} \pi^0$, 
$K^{\pm} \pi^{\mp} \pi^{\pm} \pi^{\mp}$,
$K^{\pm} \pi^{\mp} \pi^{\pm} \pi^{\mp} \pi^0$, 
$K^{\pm} \pi^{\mp} \pi^{\pm} \pi^{\mp} \pi^{\pm}$,
$K^0_S$, 
$K^0_S \pi^0$, 
$K^0_S \pi^{\pm}$, 
$K^0_S \pi^{\pm} \pi^0$, 
$K^0_S \pi^{\mp} \pi^{\pm}$,
$K^0_S \pi^{\mp} \pi^{\pm} \pi^0$,
$K^0_S \pi^{\mp} \pi^{\pm} \pi^{\mp}$,
$K^0_S \pi^{\mp} \pi^{\pm} \pi^{\mp} \pi^0$, and 
$K^0_S \pi^{\mp} \pi^{\pm} \pi^{\mp} \pi^{\pm}$.

After forming the $B \to X_{s} \ell^{+} \ell^{-}$ candidates,
we carry out the background suppression.
The largest background sources are random combinations of dileptons with
a kaon and pions that originate from continuum $\qqbar$ ($q=u,d,s,c$)
production or from semileptonic $B$ decays.  
We reject the $\qqbar$ background with a Fisher
discriminant \cite{bib:fisher} ($\Fsfw$) based on a modified set of
Fox-Wolfram moments \cite{bib:fox-wolfram} that differentiate the event
topology, by applying a cut $\Fsfw > -1.0$.
In the semileptonic $B$ decay background, both $B$ mesons
decay into leptons or two leptons are produced from the $b\to c\to s,d$
decay chain.  We combine the missing mass ($M_{\rm miss}$) and the total
visible energy ($E_{\rm vis}$) into another Fisher discriminant ($\Fmiss$)
to reject the $B$ decay background.
The missing mass is defined as 
$M_{\rm miss}
  =\sqrt{(2E_{beam}-\sum E_i)^2-|\sum{\vec{p}_i}|^2}$,  
where $E_{beam}$ is the CM beam energy
and $E_i$ ($\vec{p}_i$) is the 
reconstructed energy (3-momentum) 
for the charged particle or the photon in the CM frame.
The sum runs over all the charged particles with a pion mass hypothesis
and all photons.
The total visible energy is also calculated using the charged 
particles and photons.
We require the selection cut of $\Fmiss > 1.2$.

We further reduce the background using $\Delta E$ 
and $\chi^{2}_{B vtx}$ with the selection criteria: 
$-0.10\GeVc < \Delta E < 0.05\GeVc$ for the electron mode
($-0.05\GeVc < \Delta E < 0.05\GeVc$ for the muon mode), and 
$\chi^{2}_{B vtx}/NDF < 10.0 $.
Here, $\chi^{2}_{B vtx}$ is the fitted chi squared of the $B$ vertex
constructed from the charged daughter particles
except for the $K^0_S$ daughter tracks.
 We reject candidates with the $\Xs$ invariant mass greater than 
2.0~$\GeVcc$. This condition removes a large fraction of the combinatorial 
background while retaining 99\% of the signal.

At this stage, there is an average of 1.6 $B$ candidates 
per event in the signal simulation. 
We retain only the $B$ candidate with the largest signal likelihood. 
We select the following six variables as the background-discrimination 
variables,  
and calculate the likelihood functions based on the 
distributions of the variables:
$\DeltaE$, 
$\DeltaE^{ROE}$, 
$\chi^{2}_{B vtx}$,
$\cos\theta_B$,  
$\Fsfw$, and
$\Fmiss$.
The energy difference $\DeltaE^{ROE}$ is formed by combining all charged
tracks and neutral calorimeter clusters not included in the $B$
candidate, and 
$\cos\theta_B$ is the cosine of the
$B$ flight direction with respect to the $e^-$ beam
direction in the CM frame.

The signal probability density functions (PDFs) are determined 
by applying fits to each distribution 
for signal MC events.
For $\DeltaE$ and $\chi^{2}_{B vtx}$, 
we use the real charmonium-veto-event distributions 
to determine the PDFs, because we observe some discrepancy between 
the signal MC and the real charmonium-veto events.
In the charmonium-veto-event distributions, we subtract
the background shapes obtained by the $\Mbc$ side-band region.
The normalization of the subtracted background events
is determined by the number of background events in 
the $\Mbc$ signal region, which is estimated by 
fitting to the $\Mbc$ distribution with
a Gaussian (signal) and an ARGUS~\cite{ARGUS} (BG) function.
For $\DeltaE$, we use distinct PDFs
for the electron and muon modes.
For the background PDFs, we use background MC events.

The variables $\DeltaE$, $\DeltaE^{ROE}$, and $\Fmiss$
are effective at rejecting $\BB$ background, especially
for events with two semileptonic decays, which have large missing energy. 
For continuum suppression, the event-shape variable 
$\Fsfw$ and $\cos\theta_B$ are useful.
$\chi^{2}_{B vtx}$ is effective to reject the random combinatorial background 
in the high multiplicity modes.


We then calculate
the likelihoods $\calL_{S,B} = \prod_{ i = 1 }^{6} p_{S,B}^{i}$
where $p_{S,B}^{i}$  are the 
PDFs for the background-discrimination variable $i$ 
for the signal ($S$) and the background ($B$), respectively.
Then we obtain the final discriminating variable,
the likelihood ratio $\cal R =\calL_S/(\calL_S+\calL_B)$. 
Only the $B$~candidate with the largest ${\cal R}$
value is retained.
We find that the correct candidate is reconstructed in 84\% of the events.

\begin{figure}
\begin{center}
\includegraphics[width=14cm]{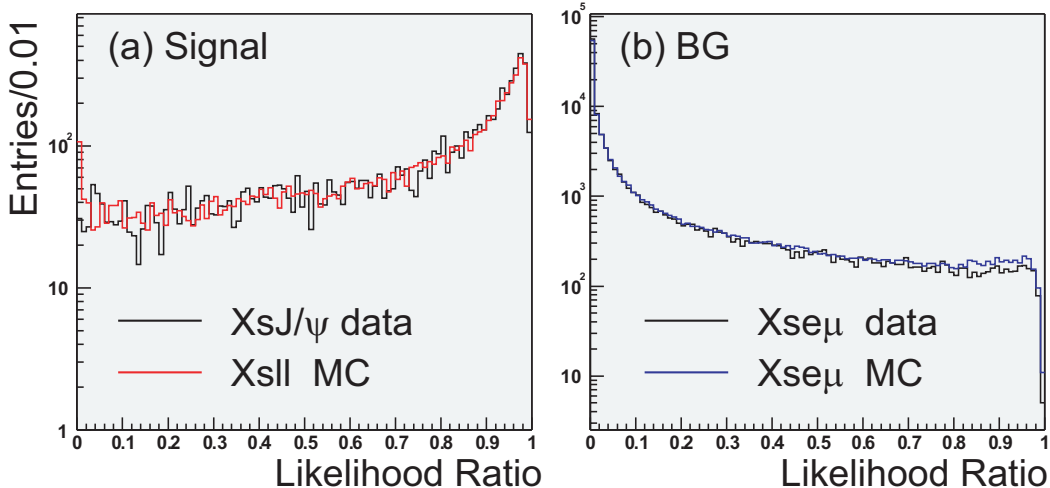}
\caption{
The likelihood ratios for (a) signal and (b) background
events. The red (blue) and black histograms in the signal (background)
likelihood distributions show signal MC (MC $\BtoXsemu$ events) 
and the real charmonium veto sample (real $\BtoXsemu$ events), respectively.
}
\label{fig:LR_varidation}
\end{center}
\end{figure}

In order to check the obtained likelihood ratio, 
we compare the real charmonium-veto events and 
signal MC for the signal likelihood ratio, and 
the real and background MC $\BtoXsemu$ events 
that are selected using the nominal selection criteria but
requiring that the two leptons have different flavor,
for the background likelihood ratio.
Figures~\ref{fig:LR_varidation}(a) and (b) show the 
likelihood ratios for the signal and background events, respectively.
We observe good agreement in both figures.

The final suppression of the combinatorial background is
achieved with the likelihood ratio ${\cal R}$.
Using the simulation, the cut on ${\cal R}$ is optimized to
maximize the statistical significance of the signal.
This optimization is performed 
in the two $m(X_{s})$
regions $m(X_{s}) < 1.1\GeVcc$, and
$1.1\GeVcc < m(X_{s}) < 2.0\GeVcc$, 
resulting in the cuts 
${\cal R} > 0.3$ and $0.9$.

After applying all selection criteria, a sample of
155 $\BtoXsee$ and 112 $\BtoXsmumu$ candidates remains
in the signal $\Mbc$ region.
Here we define the signal $\Mbc$ region as $5.27\GeVcc < M_{bc} < 5.29\GeVcc$.
According to the simulation, the background remaining
at this stage of the analysis consists
mostly of $\BB$ events (80\% and 72\% of the total
background in the electron
and muon channels, respectively). 
Using the signal MC simulation, 
the probability to select the correctly reconstructed 
candidate is estimated to be 91\%.

\section{Maximum likelihood fit}
\label{sec:Fit}
We perform an extended, unbinned maximum likelihood fit
to the $M_{bc}$ distribution 
in the region $M_{bc} > 5.2\GeVcc$ to extract the signal yield as well as
the shape and yield of the combinatorial background.
The likelihood function ${\cal L}$ is expressed as:
\[
{\cal L} = \frac{  e^{-(N_{sig} + N_{peak} + N_{total\_BG})}} { N!}
           \prod_{ i = 1 }^{N} 
            [(N_{sig} + N_{peak}) {\cal P}^{sig}_{i}
             + \sum_{BG} N_{BG} {\cal P}^{BG}_{i}
	    ] 
\]
\[
N_{total\_BG} = N_{pc} + N_{cf} + N_{comb}
\]
\[
\sum_{BG} N_{BG} {\cal P}^{BG}_{i} = 	
    N_{pc} {\cal P}^{pc}_{i} 
  + N_{cf} {\cal P}^{cf}_{i} 
  + N_{comb} {\cal P}^{comb}_{i} 
\]
where $N$ and $i$ denote the total number and index of candidate events,
respectively.
$N_{sig}$, $N_{peak}$, $N_{pc}$, $N_{cf}$, 
and $N_{comb}$ represent the yields of 
the signal, peaking background, combinatorial background from
peaking background, cross-feed events from
the mis-reconstructed $\BtoXsll$ decays, 
and combinatorial background,
respectively. 
${\cal P}^{sig}_{i}$ is the signal PDF.
We use the same PDF for signal and peaking background events.
${\cal P}^{pc}_{i}$, 
${\cal P}^{cf}_{i}$, and 
${\cal P}^{comb}_{i}$ 
are the background component PDFs 
for combinatorial background from peaking background, cross feed, 
and combinatorial background, respectively.

The signal PDF ${\cal P}^{sig}_{i}$ is described by a Gaussian
for $\mu^+ \mu^-$ modes as well as for $e^+ e^-$ modes, since the 
Bremsstrahlung recovery and selection procedure for $e^+ e^-$ modes 
lead to a negligible radiative tail in the $M_{bc}$ distribution.
The Gaussian shape parameters, the mean and the resolution $\sigma$ 
values are determined  
from fits to the sum of a Gaussian and an ARGUS function 
for the real charmonium-veto data sample.
The fits results in a signal $M_{bc}$ mean and resolution, respectively, of 
$m_{sig} = 5279.31 \pm 0.05\MeVcc$ and 
$\sigma_{sig} = 2.62 \pm 0.04\MeVcc$ for the $e^+ e^-$ modes,
and
$m_{sig} = 5279.03 \pm 0.04\MeVcc$ and
$\sigma_{sig} = 2.53 \pm 0.04\MeVcc$ for the $\mu^{+} \mu^{-}$ modes.
In the simulation,
the Gaussian fit results for the $M_{bc}$ distributions for correctly
reconstructed signal are in agreement with the shape parameters
extracted from the fits to the charmonium-veto sample.
The signal yield $N_{sig}$ is a free parameter in the likelihood fit.
 
The charmonium peaking background is estimated from the simulation
to be $1.20 \pm 0.28$ events in the electron modes and
$1.33 \pm 0.21$ events in the muon modes.
The charmonium peaking background PDF is the same as that 
for signal since the signal PDF is extracted 
from the charmonium-veto data sample.

\begin{figure}
\begin{center}
\includegraphics[width=12cm]{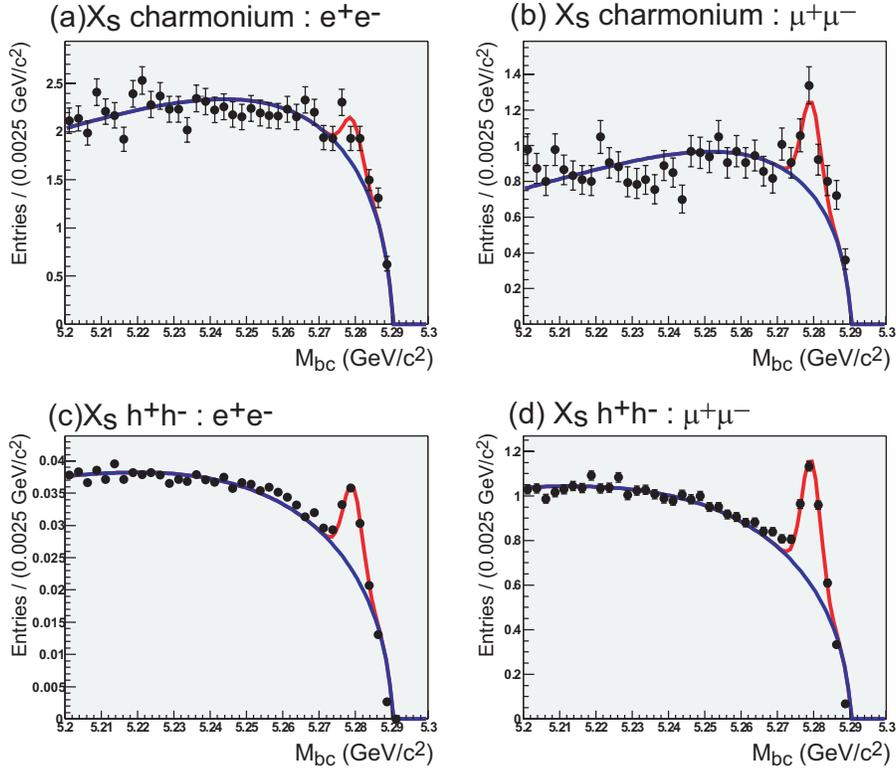}
\caption{
$\Mbc$ distributions 
for MC charmonium events for (a) electron and (b) muon modes, 
and real $B \rightarrow X_s h^+h^-$ events for (c) electron 
and (d) muon modes.
}
\label{fig:peaking}
\end{center}
\end{figure}

The size and shape of the hadronic peaking $\BB$ background component arising 
from $B \to D^{(*)} n \pi$~$(n > 0)$ decays with misidentification 
of two charged pions as leptons
are derived directly from the real data 
by performing the analysis without the lepton identification requirements. 
By fitting the $M_{bc}$ distribution to the sum of 
Gaussian and ARGUS functions, we get the mean and the resolution of
$m_{peak} = 5279.16 \pm 0.04\MeVcc$ and
$\sigma_{peak} = 2.60 \pm 0.02\MeVcc$.
Taking the $\pi$ to $\ell$ misidentification rates into
account, the remaining hadronic peaking background 
is estimated to be 
$N_{h-peak} = 0.03 \pm 0.001$ events for the $e^+ e^-$ modes
and 
$N_{h-peak} = 1.78 \pm 0.05$ events for the $\mu^{+} \mu^{-}$ modes. 
Here we use the momentum and polar-angle dependent 
misidentification rate.
The average misidentification rates for electrons and 
muons are 0.08\% and 0.92\%, respectively.
In the likelihood fit, we also use the same signal PDF for the
hadronic peaking-background PDF, 
because the fitted Gaussian shape-parameter values are 
consistent with those of the charmonium-veto data sample, 
and $N_{peak}$ is fixed to the estimated values.
Figures~\ref{fig:peaking}(a) and (b) show the $\Mbc$ distributions 
for the MC charmonium events for the electron and muon modes, and 
(c) and (d) show the $\Mbc$ distributions for real 
$B \rightarrow X_s h^+h^-$ events for the electron and muon modes, respectively.

The background component PDFs, 
${\cal P}^{pc}_{i}$, 
${\cal P}^{cf}_{i}$, and 
${\cal P}^{comb}_{i}$ 
are given by an ARGUS shape.
They describe the
combinatorial contribution from peaking background events,
from cross feed events, and
from continuum events and $\BB$ events, respectively.
The ARGUS cutoff is determined by the 
beam energy in the $\Upsilon(4S)$ rest frame, $E_{beam} = 5.290\GeV$.
The values of the ARGUS shape parameter for each background component
are determined 
from the peaking background $M_{bc}$ distribution shown in 
Fig.\ref{fig:peaking} (${\cal P}^{pc}_{i}$), 
from incorrectly reconstructed signal MC events
(${\cal P}^{cf}_{i}$), 
and from the fit to the real $\BtoXsemu$ events
selected using the nominal selection criteria but
requiring that the two leptons 
have different flavor(${\cal P}^{comb}_{i}$).
We fix the three ARGUS shape parameters, 
$N_{pc}$, and $N_{cf}$.
The yield $N_{comb}$ is 
taken as a free parameter in the likelihood fit. 

\section{Results}
\label{sec:Results}

\begin{figure}[!htb]
\begin{center}
\includegraphics[width=16cm]{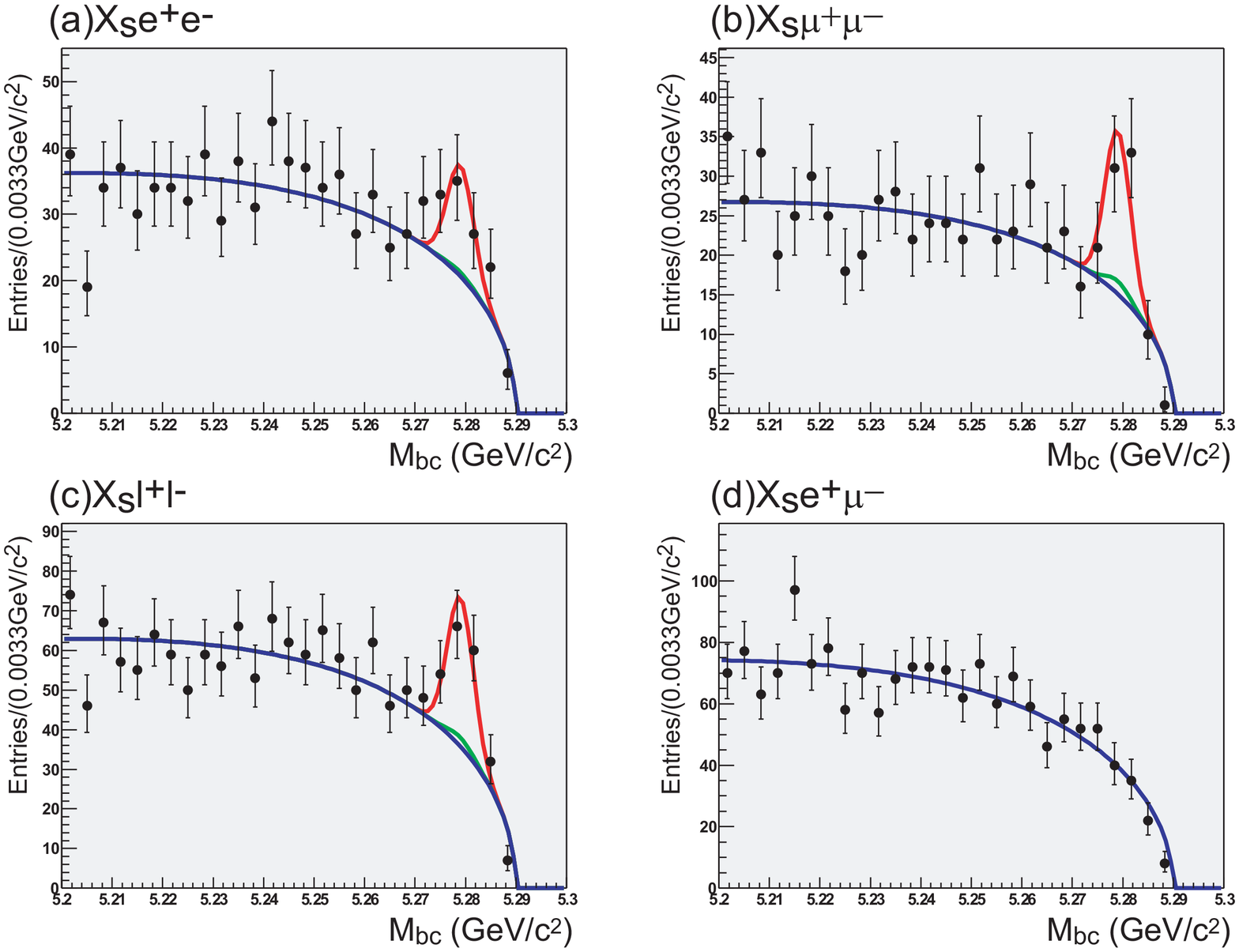}
\caption{Distributions of $M_{bc}$ for selected
(a) $\BtoXsee$, (b) $\BtoXsmumu$, (c) $\BtoXsll$ ($\ell = e, \mu$), 
and (d) $\BtoXsemu$ candidates.
The red lines represent the result of the fits,
and the green and blue lines represent 
the peaking and combinatorial background 
components under the signal peaks, respectively.}
\label{fig:fitfig}
\end{center}
\end{figure}

Using the fit parameterizations described above, 
we fit the $M_{bc}$ distributions for the selected
$\BtoXsee$ and $\BtoXsmumu$ candidates separately 
and obtain the results shown in Figure~\ref{fig:fitfig}.
The fit results are summarized in Table~\ref{tab:fitresult}.
The statistical significance is
${\cal S} = \sqrt{2 (\ln{\cal L}_{max} - \ln{\cal L}^{0}_{max})}$,
where ${\cal L}_{max}$ represents the maximum likelihood for the
fit and ${\cal L}^{0}_{max}$ denotes the maximum likelihood for a
different fit when the signal yield is fixed at $N_{sig} = 0$.
The $\BtoXsll$ signal yield presented in Table~\ref{tab:fitresult} is
the sum of the $\BtoXsee$ and $\BtoXsmumu$ signal yields.
A separate fit to the combined electron and muon channels
gives a comparable result.
Figure~\ref{fig:fitfig}(d) shows the $M_{bc}$ distribution for $\BtoXsemu$ 
candidates. 
Applying the ARGUS fit to the $M_{bc}$ distribution, 
there is no evidence for a peaking background as expected.

\begin{table}[b]
 \caption{Results of the fit to the data: number of the signal candidates
 in the signal box, obtained signal yield,
  peaking backgrounds (fixed in the fit),
  and statistical significance.}
 \begin{center}
 \begin{ruledtabular}
 \begin{tabular}{lcccc}
   Mode & 
       Candidates & $N_{sig}$ & $N_{peak}$ & Signif.\\
  \hline
  $X_s\: e^+ e^-$ &    
       155 & $31.8 \pm 10.2$ & $1.24 \pm 0.28$ & \SigBtoXsee \\
  $X_s\: \mu^+\mu^-$ & 
       112 & $36.3 \pm ~9.3$ & $3.11 \pm 0.22$ & \SigBtoXsmumu \\
  $X_s\: \ell^+\ell^-$ &
       267 & $68.4 \pm 13.8$ & $4.35 \pm 0.36$ & \SigBtoXsll \\
 \end{tabular}
 \end{ruledtabular}
 \end{center}
 \label{tab:fitresult}
\end{table}

\begin{figure}
\begin{center}
\includegraphics[width=16cm]{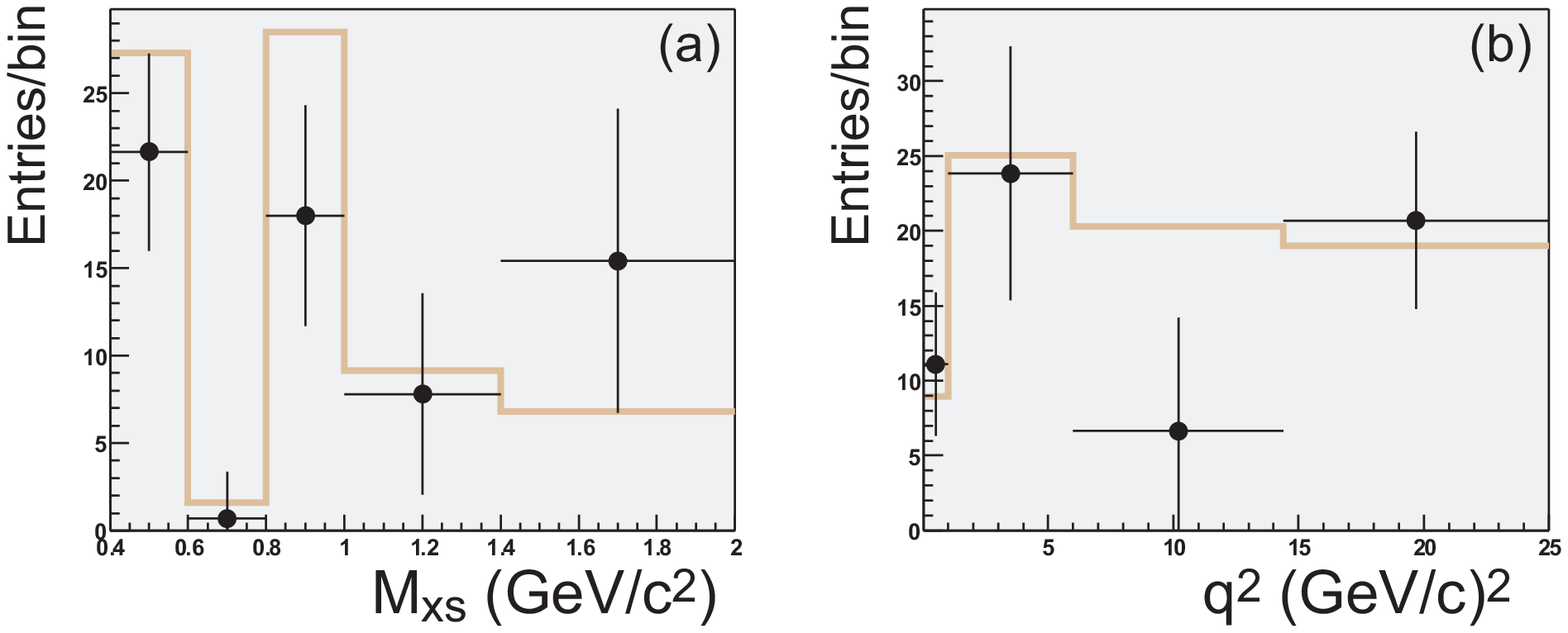}
\caption{Distributions of number of signal events as a function
 of (a) hadronic mass $\Mxs$ and (b) $q^2 \equiv \Mll^2$ 
for electron and muon channels combined
 for data (points) and Monte Carlo signal (histogram).
 The vertical error bars represent statistical errors only.}
\label{fig:gra}
\end{center}
\end{figure}

Figures~\ref{fig:gra}(a) and (b) show the distributions for the 
hadronic mass $\Mxs$
and $q^2 \equiv \Mll^2$ for the electron and muon channels combined,
but they are obtained by performing the nominal likelihood fit
in separate $\Mxs$ and $q^2$ regions.
Figure~\ref{fig:gra}(a) indicates that the observed signal includes contributions
from final states across a range of hadronic masses, including
hadronic systems with a mass above that of the $K^\ast(892)$. 

The branching fraction ${\cal B}$ for the signal is calculated from
\begin{equation}
  {\cal B} = \frac{ N_{sig} }{ 2 N_{\BB}~ \epsilon },
\end{equation}
where $N_{\BB} = (152.0 \pm 0.7) \times 10^{6}$ is the number
of $\BB$ pairs produced in $140.0~\fbinv$ and
$\epsilon$ is the signal efficiency.

\section{Systematic uncertainties}
\label{sec:Systematics}

Systematic uncertainties are of two different
types: those that affect the extraction of the number of signal events
and those that affect the calculation of the branching fraction.
The systematic uncertainties are summarized in Table~\ref{tab_systematics}.

\begin{table}
\caption{Summary of fractional systematic uncertainties (in \%).
 The uncertainties in extracting the signal are presented first and
 those related to the signal efficiency and $\BB$ counting are
 presented second.}
\begin{center}
\begin{tabular}{lrr}
\hline \hline
Source    	        & $X_s\: e^+e^-$  & $X_s\: \mu^+\mu^-$   \\ \hline\hline
Signal shape                   & $\pm 1.4$       & $\pm 0.5$    \\
BG shape                       & $\pm 7.8$       & $\pm 4.7$    \\
Peaking background statistics  & $\pm 0.9$       & $\pm 0.6$    \\ 
Peaking background shape       & $\pm 4.3$       & $\pm 2.1$    \\ 
Cross-feed events              & $\pm 4.1$       & $\pm 2.2$    \\ \hline\hline
~~~~Signal yield total         & $\pm 9.9$       & $\pm 5.7$    \\ \hline\hline
Tracking efficiency            & $\pm 3.5$       & $\pm 3.5$    \\
Lepton identification efficiency    & $\pm 1.0$       & $\pm 2.4$    \\
Kaon identification efficiency      & $\pm 0.8$       & $\pm 0.8$    \\
$\pi^\pm$ identification efficiency & $\pm 0.6$       & $\pm 0.5$    \\
$K^0_S$ efficiency             & $\pm 0.7$          & $\pm 0.8$    \\
$\pi^0$ efficiency          & $\pm 0.3$          & $\pm 0.3$    \\
$\cal R$ cut efficiency     & $\pm 5.4$          & $\pm 1.5$    \\ \hline
~~~~Detector model subtotal & $\pm 6.6$          & $\pm 4.7$    \\ \hline
Fermi motion model          & $^{+ 6.5}_{-2.4}$  &$^{+ 6.1}_{-2.3}$ \\
${\cal B}(\BtoKll)$         & $\pm 6.4$          & $\pm 6.8$    \\
${\cal B}(\BtoKstarll)$     & $\pm 7.0$          & $\pm 7.8$    \\
$K^\ast$--$X_s$ transition  & $\pm 4.5$          & $\pm 4.7$    \\
Hadronization               & $\pm 8.5$          & $\pm 8.2$    \\
Missing modes               & $\pm 4.5$          & $\pm 4.4$    \\ \hline
~~~~Signal model subtotal   & $^{+15.6}_{-14.4}$ & $^{+15.9}_{-14.9}$ \\ \hline
Monte Carlo statistics      & $\pm 1.6$          & $\pm 1.5$    \\
$\BB$ counting               & $\pm 0.5$          & $\pm 0.5$    \\
\hline\hline
~~~~Efficiency and $N_{\BB}$ total 
                            & $^{+17.0}_{-15.9}$ & $^{+16.6}_{-15.7}$   \\
\hline\hline
~~~~Total                   & $^{+19.7}_{-18.8}$ & $^{+17.6}_{-16.7}$   \\
\hline\hline
\end{tabular}
\end{center}
\label{tab_systematics}
\end{table}

Uncertainties affecting the extraction of the signal yield
are evaluated by varying the signal Gaussian parameters (mean and width)
and the background shape parameter within $\pm$1$\sigma$ of the
measured values from the charmonium veto data (signal) and 
the real $\BtoXsemu$ events (background).

We estimate the uncertainties in the peaking-background shape 
and cross-feed events by comparing the signal yields 
obtained with and without the corresponding PDFs 
in the unbinned maximum likelihood fit to $M_{bc}$.

Uncertainties affecting the signal efficiency originate from
the detector modeling, from the simulation of signal decays,
and from the estimate of the number of $B$ mesons in the sample.
By far the largest component is that due to the
simulation of signal decays, discussed in detail below.

The detector modeling uncertainty is sensitive to the
following uncertainties determined from the data:
the uncertainty in the tracking efficiency of $1.0$\% per track;
the uncertainty in the charged-particle
identification efficiency of $0.5$\% per electron, $1.2$\% per muon,
$1.0$\% per kaon and $0.8$\% per pion; 
and the uncertainty in the reconstruction
efficiency of $4.5$\% per $K^0_S$ and $3.3$\% per $\pi^0$.
The efficiency of the likelihood ratio cut, which suppresses combinatorial
background, is checked with the charmonium-veto sample and the level
of discrepancy with the simulation is taken as the corresponding uncertainty.

The dominant source of uncertainty arises from modeling the signal decays.
Parameters of the Fermi motion model are varied in accordance with
measurements of hadronic moments in semileptonic $B$
decays~\cite{CLEO_moments} and the photon spectrum in inclusive $B \to
X_s\:\gamma$ decays~\cite{CLEO_bsgamma}. 
The fractions of exclusive $\BtoKll$ and $\BtoKstarll$ decays
are varied according to experimental~\cite{Belle_Kstll,BaBar_Kstll} and
theoretical uncertainties~\cite{Ali02}, respectively.
The transition point in $m(X_s)$ between pure $K^\ast \ell^+ \ell^-$
and non-resonant $X_s\:\ell^+\ell^-$ final states is varied
by $\pm 0.1\GeVcc$.

The non-resonant Monte Carlo event generator relies on JETSET to
fragment and hadronize the system consisting of a final
state $s$ quark and a spectator quark from the $B$ meson.
Since the signal efficiencies depend strongly on the
particle content of the final state, uncertainties in the
number of charged and neutral pions and in the 
number of charged and neutral kaons translate
into a significant uncertainty in the signal efficiency
(for $m(X_s) > 1.1\GeVcc$).

The ratio between the generator yield for decay modes containing
a $K^0_S$ and that for modes containing a charged kaon
is varied according to $0.50 \pm 0.11$,
to allow for isospin violation in the decay chain.
The ratio between the generator yield for
decay modes containing one $\pi^0$ meson and that for modes
containing none is varied according to $1.0 \pm 0.22$.
Uncertainties in the ratios are set by the
level of discrepancy between $\BtoJpsiX$ real data 
and $\BtoXsll$ Monte Carlo event sample.

The 18 modes selected in this analysis only capture about
53\% of the full set of final states.
Approximately 60\% of the missing modes are due to
final states with a $K^0_L$ meson and their contribution can be determined
from the $K^0_S$ modes.
However, we need to account for the uncertainty in the fraction of modes
with too many pions or kaons (two extra kaons may be
produced via $s \bar{s}$ popping),
as well as for modes with photons
that do not originate from $\pi^0$ decays but rather
from $\eta$, $\eta^\prime$, etc.
For final states with $m(X_s) > 1.1\GeVcc$, we vary
these fractions by 
$\pm 5\%$ per $\pi^0$, 
$\pm 20\%$ for $\eta$, 
$\pm 30\%$ for $N_{\pi}>5$, and
$\pm 50\%$ for $\eta^{\prime}$ and others.

Including systematic uncertainties, the measured branching fractions
for $m(\ell^{+} \ell^{-}) > 0.2\GeVcc$ are
\begin{eqnarray}
 {\cal B}(\BtoXsee)   & = & \left(\BrBtoXsee\right)
 \times 10^{-6}, \\
 {\cal B}(\BtoXsmumu) & = & \left(\BrBtoXsmumu\right)
 \times 10^{-6}, \\
 {\cal B}(\BtoXsll)   & = & \left(\BrBtoXsll\right)
 \times 10^{-6},
\end{eqnarray}
where the first error is statistical and the second error is systematic.
The combined $\BtoXsll$ branching fraction is the weighted average of
the branching fractions for the electron and muon channels,
where we assume the individual branching fractions to be equal 
for $m(\ell^{+} \ell^{-}) > 0.2\GeVcc$.
Table~\ref{tab_results} summarizes the results of the analysis
and lists both the statistical and systematic errors in the signal
yields, the signal efficiencies and the branching fractions.

\begin{table}
 \caption{Summary of results: signal yield ($N_{sig}$),
  statistical significance (Signif.), efficiency ($\epsilon$) 
  and branching fraction ($\cal B$).
  In the case of the signal yield and the branching fraction,
  the first error is statistical and the second error is systematic.
  In the case of the signal efficiency, the first error corresponds
  to uncertainties in detector modeling, $\BB$ counting, and Monte Carlo
  statistics, whereas the second error corresponds to the uncertainties
  in the signal model.}
 \baselineskip=28pt
 \begin{center}
 \begin{tabular}{lcccc}
  \hline \hline 
  Mode 
    & $N_{sig}$ & Signif. & $\epsilon$ (\%) & ${\cal B}~(\times 10^{-6})$ \\
  \hline\hline
  $X_s\: e^+ e^-$
    & $31.8 \pm 10.2 \pm 3.1$ & \SigBtoXsee
    & $\EffBtoXsee $ & $\BrBtoXsee$ \\
  $X_s\: \mu^+\mu^-$
    & $36.3 \pm ~9.3 \pm 2.1$ & \SigBtoXsmumu
    & $\EffBtoXsmumu $ & $\BrBtoXsmumu$ \\
  $X_s\: \ell^+\ell^-$
    & $68.4 \pm 13.8 \pm 5.0$ & \SigBtoXsll
    & $\EffBtoXsll $ & $\BrBtoXsll$ \\
  \hline
 \end{tabular}
 \end{center}
 \label{tab_results}
\end{table}

\begin{figure}
\begin{center}
\includegraphics[width=16cm]{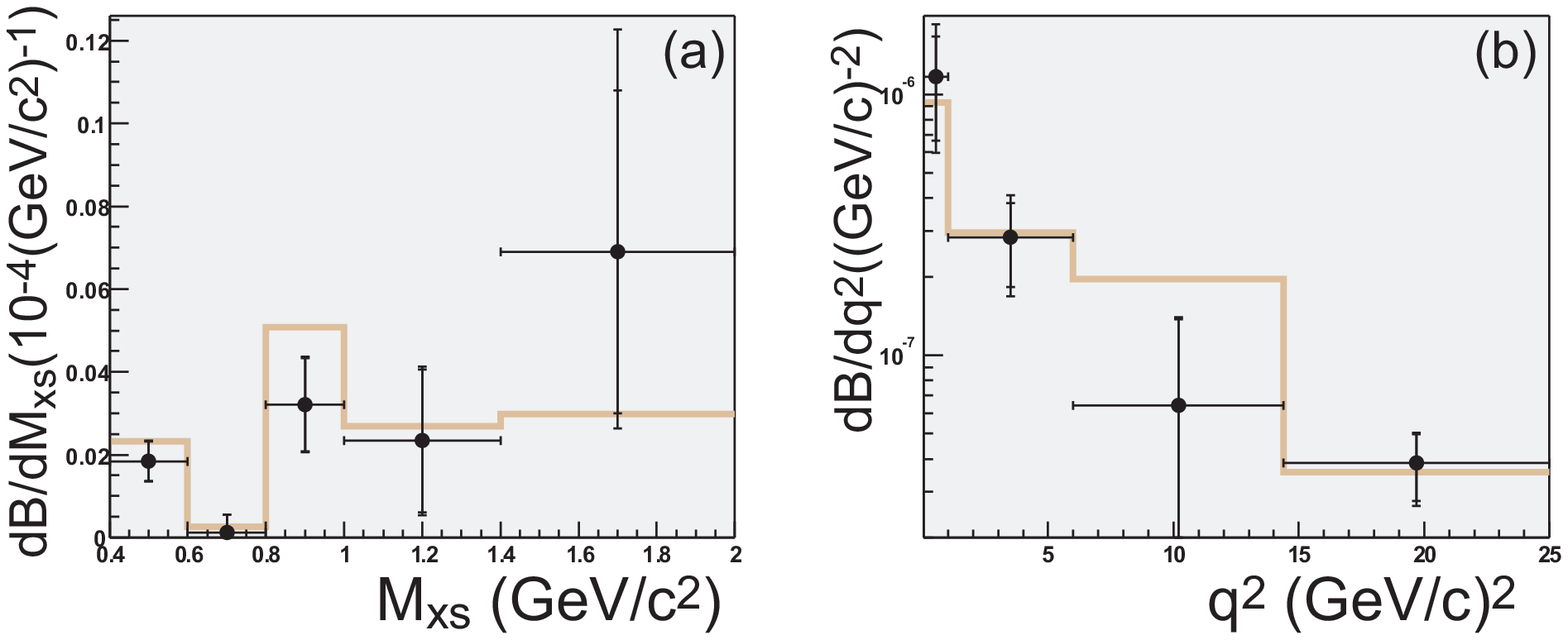}
\caption{Differential branching fraction 
as a function
 of (a) hadronic mass $\Mxs$ and (b) $q^2 \equiv \Mll^2$ 
for electron and muon channels combined
 for data (points) and Monte Carlo signal (histogram).
The outer (inner) error bars represent the total (statistical)
 errors.}
\label{fig:gra_br}
\end{center}
\end{figure}

The branching fractions for each $\Mxs$ and $q^2$ bin 
are also measured, and summarized in Table~\ref{tab_diffBR}.
Figures~\ref{fig:gra_br}(a) and (b) show the distributions of
the differential branching fractions as a function
 of (a) hadronic mass $\Mxs$ and (b) $q^2 \equiv \Mll^2$ 
for electron and muon channels combined.

\begin{table}
 \caption{
  Branching fractions ($\Br$) for each bin of $\Mxs$ and $q^2$. 
  The first and second errors are statistical 
  and systematic, respectively. 
}
 \baselineskip=28pt
 \begin{center}
 \begin{tabular}{lcclc}
  \hline \hline 
   bin & ${\cal B}~(\times 10^{-7})$ & \,\,\,\,\, &
   bin & ${\cal B}~(\times 10^{-7})$ \\
  \hline
$\Mxs (\GeVcc)$ &  & & $q^2 (\GeVc)^{2}$ & \\
$[0.4,0.6]$ & $3.75 \pm 0.96 ^{+0.23}_{-0.23}$ & &
$[0.04,1.0]$
     & $11.34 \pm 4.83 ^{+4.51}_{-2.57}$ \\

$[0.6,0.8]$ & $0.36 \pm 0.88 ^{+0.08}_{-0.08}$ & &
$[1.0,6.0]$
     & $14.93 \pm 5.04 ^{+3.82}_{-2.83}$ \\

$[0.8,1.0]$ & $6.65 \pm 2.25 ^{+0.51}_{-0.51}$ & &
$[6.0,14.4]$
     & $7.32 \pm 6.14 ^{+1.69}_{-1.77}$ \\

$[1.0,1.4]$ & $10.50 \pm 6.90 ^{+1.88}_{-1.97}$ & &
$[14.4,25.0]$
     & $4.18 \pm 1.17 ^{+0.49}_{-0.57}$ \\

$[1.4,2.0]$ & $46.59 \pm 23.37 ^{+22.31}_{-10.51}$ & &
     & \\

  \hline

  \hline
  \hline
 \end{tabular}
 \end{center}
 \label{tab_diffBR}
\end{table}

\section{Summary}
\label{sec:Summary}
Using a sample of $152 \times 10^{6}$
$\Upsilon(4S) \to B\overline{B}$ events,
we measure the branching fraction for the rare decay 
$\BtoXsll$, where $\ell = e$ or $\mu$ 
and $\Xs$ is a hadronic system is semi-inclusively 
reconstructed using 18 different hadronic states (with up to four pions).
For $m(\ell^{+} \ell^{-}) > 0.2\GeVcc$, we observe a signal of
$68.4 \pm 13.8(stat) \pm  5.0(syst)$ events
and obtain a branching fraction of
\[
   \Br(\BtoXsll)=(\BrBtoXsllFull)\times10^{-6},
\]
with a statistical significance of $\SigBtoXsll\:\sigma$.

This result is consistent with the recent prediction
by Ali {\em et al.}~\cite{Ali02}, 
our previous inclusive $\BtoXsll$ measurement~\cite{Belle_sll}, 
and that of the BaBar collaboration~\cite{BaBar_sll}, 
within errors.


\section*{Acknowledgments}
\label{sec:Acknowledgments}

The authors wish to thank Gudrun~Hiller, Tobias~Hurth and 
Gino~Ishidori for their helpful suggestions.
We have also benefited from suggestions by Stephane Willocq 
regarding event generation with EVTGEN.
We are grateful for the KEKB accelerator group for their excellent
operation of the KEKB accelerator.
We acknowledge support from the Ministry of Education,
Culture, Sports, Science, and Technology of Japan
and the Japan Society for the Promotion of Science;
the Australian Research Council
and the Australian Department of Industry, Science and Resources;
the National Science Foundation of China under contract No.~10175071;
the Department of Science and Technology of India;
the BK21 program of the Ministry of Education of Korea
and the CHEP SRC program of the Korea Science and Engineering Foundation;
the Polish State Committee for Scientific Research
under contract No.~2P03B 17017;
the Ministry of Science and Technology of the Russian Federation;
the Ministry of Education, Science and Sport of the Republic of Slovenia;
the National Science Council and the Ministry of Education of Taiwan;
and the U.S.\ Department of Energy.


\end{document}